# TRIANGLES AS BASIS TO DETECT COMMUNITIES: AN APPLICATION TO TWITTER'S NETWORK

Y. Abdelsadek[1, 2*], K. Chelghoum[1], F. Herrmann[1], I. Kacem[1] and B. Otjacques[2]

[1]Laboratoire de Conception, Optimisation et Modélisation des Systèmes (LCOMS EA 7306)
University of Lorraine, Metz, France
{kamel.chelghoum; francine.herrmann; imed.kacem}@univ-lorraine.fr

[2]e-Science Research Unit, Environmental Research and Innovation
Luxembourg Institute of Science and Technology
Belvaux, Luxembourg
{youcef.abdelsadek; benoit.otjacques}@list.lu

## ABSTRACT

Nowadays, the interest given by the scientific community to the investigation of the data generated by social networks is increasing as much as the exponential increasing of social network data. The data structure complexity is one among the snags, which slowdown their understanding. On the other hand, community detection in social networks helps the analyzers to reveal the structure and the underlying semantic within communities. In this paper we propose an interactive visualization approach relying on our application *NLCOMS*, which uses synchronous and related views for graph and community visualization. Additionally, we present our algorithm for community detection in networks. A computation study is conducted on instances generated with the LFR [9]-[10] benchmark. Finally, in order to assess our approach on real-world data, we consider the data of the ANR-Info-RSN project. The latter addresses community detection in Twitter.

**Keywords**: Social network analysis, weighted graphs, interactive visualization, community detection algorithm, visualization tool.

## 1    INTRODUCTION:

The graphs are well-known to be suitable for representing relational data in various disciplines, where the nodes and the edges, respectively, represent the entities and the relationship between these entities. For example, in sociology, graphs represent friendship in a social blog or phone-call between customers of a mobile operator, like in Blondel et al. [3] or e-mail communication like in Schlitter et al. [20]. Moreover, in biology, graphs can represent interaction between proteins, Rahman and Ngom [17]. In real-world and in common cases the edges are not binary but they are weighted. For example, the weight in the e-mail network could be the number of e-mails sent between two persons. This leads to complex and unstructured graphs, which makes the investigation tasks harder. In this context, one among the objectives in this work is to explain how the information is shared on social network.

Furthermore, the objective of community detection algorithms is to reveal the semantic of the underlying network structure communities. But what is a community? A widespread definition introduces a community as a group of network's nodes having more interactions between them comparing to the other nodes. To detect such communities, in this paper we propose a community detection algorithm in weighted graphs based on the weighted maximum triangle packing to build the skeleton of communities at its first step. Thereafter, the algorithm compares the intra-community weight and inter-community weight between groups allowing dominant communities to gain in size.

---

[*] Corresponding Author: youcef.abdelsadek@univ-lorraine.fr

Community detection is the preliminary step to grasp the underlying semantic and the structural information in the network. It comes after the visualization issue. What is the most appropriate visualization for the detected communities and the underling information? As examples, Blondel et al. [3] use node-link for community depiction whereas word cloud is used in Yang et al. [22]. In this work, additionally to node-link representation, circle packing is used to visualize the detected communities. Also, synchronous and coordinated views help the expert user to build his/her own ideas about communities' characteristics.

The remaining part of this paper is organized as follows. In Section 2, we survey the related work to the community detection algorithms. In Section 3, we give some definitions and notations used in this paper and describe our algorithm. Experimental study of the proposed algorithm is conducted in Section 4. Section 5 introduces the case study and discusses the results relying on *NLCOMS*. Finally, Section 6 concludes the paper and discusses further improvements of this work.

## 2  RELATED WORK

In general, there are two major approaches for community detection in graphs. The first approach consists in computing a similarity or a distance function between each couple of nodes. Equation (1) is an example of distance function, noted as $dist_{n_i, n_j}$, where $e^w_{n_i, n_j}$ represents the positive value of the weighted edge between the $i^{th}$ and $j^{th}$ nodes and $Const$ represent a constant. Then, a clustering algorithm is applied, like in Schlitter et al. [20]:

$$dist_{n_i, n_j} = \begin{cases} 0 & , if\ n_i = n_j \\ \frac{1}{e^w_{n_i, n_j}} & , if\ e^w_{n_i, n_j} > 0 \\ Const & , if\ e^w_{n_i, n_j} = 0 \end{cases} \qquad (1)$$

The second approach relies on the graph structure. In this second approach, there are two families of algorithms. The divisive family or top-down: initially the whole graph is considered as a unique community and a divisive algorithm tries iteratively to prune the inter-community edges. As an example, we refer to the divisive algorithm of Newman and Girvan [12]. The algorithm uses the *edge betweenness centrality* to detect such inter-community edges, where the *edge betweenness centrality* metric counts the number of short-paths between all couples of nodes in which this edge belongs. The second family is the agglomerative algorithms or bottom-up: an agglomerative algorithm starts with $v$ communities, $v$ representing the number of nodes in the graph, which means that initially each node forms a community. Thereafter, the aim is to merge communities with respect to an objective function or a metric. One of the common metric used in Newman [13] and Clauset et al. [5], is the *modularity $\varphi$*, formulated by Equation (2):

$$\varphi = \frac{1}{2M} \sum_{n_i} \sum_{n_j} \left( e^w_{n_i, n_j} - \frac{WD_{n_i}\ WD_{n_j}}{2M} \right) \delta(c_{n_i}, c_{n_j}) \qquad (2)$$

Where:

- $M = \sum_{n_i} \sum_{n_j} e^w_{n_i, n_j}$,
- $c_{n_i}$ is the community of $n_i$,
- $\delta\left(c_{n_i}, c_{n_j}\right) = 1$, if $c_{n_i} = c_{n_j}$, 0 otherwise,

- $WD_{n_i}$ is the weighted graph degree of the node $n_i$ defined as follows: $WD_{n_i} = \sum_{j=1}^{v} e^{w}_{n_i, n_j}$,
- Function $\varphi$ is used to express whether the weights of edges inside communities are greater than a random edge distribution with the same properties.

The aforementioned community definition with more edges within communities than outside is still valid for the weighted context. Indeed, exchanging each weighted edge by as much edges having 1 as weight leads to a multigraph [14]. As an example of agglomerative algorithm we refer to the algorithm of Blondel et al. [3]. In the first phase of their algorithm, a merging step is achieved until a local maximum of *modularity* is reached; no merging can improve the *modularity*. In the second phase, each detected community is replaced by a meta-node. These two phases are iteratively performed until reaching a global maximum of the *modularity*. We refer to the papers of Leskovec et al. [11] and Lancichinetti et al. [8] for an evaluation of community detection algorithms.

# 3    OUR ALGORITHM

Before introducing our proposed algorithm for community detection, we give some definitions and notations which are necessary for a better understanding.

## 3.1    Definitions and Notations

Let $G = (N, E, E^w)$ be a graph where $N$, $E$ and $E^w$ denote respectively the set of nodes of size $v$, the set of edges of size $m$ and the weights of the edges. The set of communities is $Cs = \{c_1, c_2 \ldots c_k\}$, where $k \leq v$. Therefore, $WD_{c_g} = \sum WD_{n_i}, \forall n_i \in c_g$. The intra-weight $IW_{c_g}$ and the inter-weight $INW_{c_g, c_h}$, are respectively defined as follows:

- the sum of the weights of the edges within a community $c_g$: $IW_{c_g} = \sum_{i<j} (e^{w}_{n_i, n_j}), \forall n_i, \forall n_j \in c_g$
- the sum of the weights of edges between two communities $c_g$ and $c_h$ : $INW_{c_g, c_h} = \sum_{i<j} (e^{w}_{n_i, n_j}), \forall n_i \in c_g$ and $\forall n_j \in c_h$.

Two communities $c_g$ and $c_h \in Cs$ are adjacent if $INW_{c_g, c_h} \neq 0$. $|S|$ denotes the cardinality of the set $S$. What makes a community dominant is its $intra-weight$ or its $weighted\ degree$, greater are these values, more the community is dominant. The neighbourhood degree $\lambda_l$ of triangle $l$ represents the number of triangles that overlap $l$.

The objective of the unweighted Maximum Triangle Packing problem (MTP for short) is to find the largest collection of pairwise node-disjoint triangles (i.e., cliques of size 3) of a graph among all possible triangles $t = 1, 2, \ldots, T$, (see Abdelsadek et al. [1] for algorithms addressing the MTP problem). A weighted version of the MTP can be formulated as in Model (3), where $x$ is the decision variables' vector, $t^l$ is the weight of triangle $l$ obtained by summing the weights of its edges and $B_{i\,l} \in \{0, 1\}^{v \times T}$ is the node-triangle belonging matrix. We point out that there exists another formulation of the weighted MTP in Chen et al. [4].

$$Maximize\ \sum_{l=1}^{T} t^l \times x_l$$
$$s.t.\ \sum_{l=1}^{T} B_{i\,l} \times x_l \leq 1, \qquad \forall i \in \{1 \ldots v\} \qquad\qquad (3)$$
$$x_l \in \{0, 1\}, \qquad\qquad \forall l \in \{1 \ldots T\}$$

For instance, let us consider the graph in Figure 1. In the latter a greedy solution for the weighted MTP can be obtained by selecting triangles in a decreasing order with respect to their $t^l$ value. Thus, we obtain the following result: $\{(n_3, n_4, n_9), (n_7, n_8, n_{10}), (n_{11}, n_{13}, n_{16}), (n_{12}, n_{17}, n_{18})\} = 8 + 8 + 8 + 8 = 32$. The heuristic that we propose for the weighted MTP problem consists in selecting the triangles in a decreasing order with respect to the value returned by the *evaluation* function $E$ defined in (4) while the disjunctive constraint is met.

$$E\ (l) = \begin{cases} t^w / \lambda_l & , \ if \ \lambda_l \neq 0 \\ t^w & , \ if \ \lambda_l = 0 \end{cases} \qquad\qquad \textbf{(4)}$$

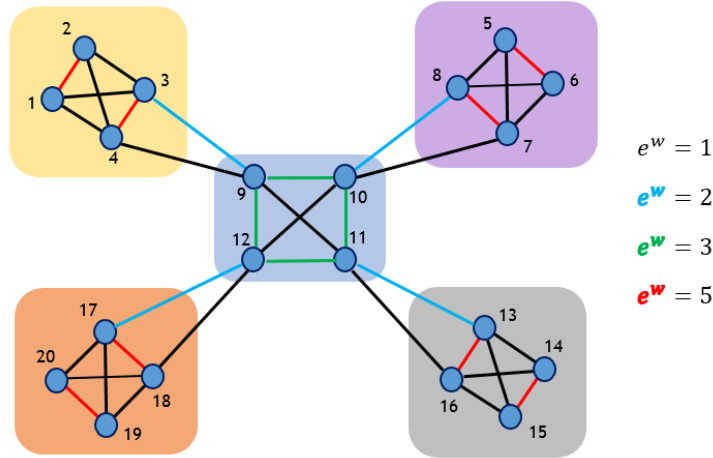

**Figure 1: Graph example, with $e^w = 1$ for the black edges, $e^w = 2$ for the blue edges, $e^w = 3$ for the green edges and $e^w = 5$ for the red edges.**

In Figure 1, $\{(n_1, n_3, n_4), (n_6, n_7, n_8), (n_9, n_{10}, n_{11}), (n_{13}, n_{14}, n_{16}), (n_{17}, n_{18}, n_{20})\} = 7 + 7 + 7 + 7 + 7 = 35$ is returned as solution for the weighted MTP problem. This solution is optimal but is not the only optimal solution for this example. The following are the *evaluation* functions of the obtained solution:

$E\ (n_1, n_3, n_4) = 7/4 = 1.75;$
$E\ (n_6, n_7, n_8) = E\ (n_{13}, n_{14}, n_{16}) = E\ (n_{17}, n_{18}, n_{20}) = 7/4 = 1.75;$
$E\ (n_9, n_{10}, n_{11}) = 7/6 \cong 1.17;$
$E\ (n_3, n_4, n_9) = 8/8 = 1;$
$E\ (n_7, n_8, n_{10}) = E\ (n_{11}, n_{13}, n_{16}) = E\ (n_{12}, n_{17}, n_{18}) = 8/8 = 1.$

The motivation behind this is to select triangles as a skeleton of the communities' structure. Topologically speaking, triangles are the smallest group of interconnected nodes after edges and they are considered in various networks metrics. As an example, we refer to the *global clustering coefficient* of Opsahl and Panzarasa [16], the *cohesion* of Friggeri et al. [6] and the *3-cycle cut ratio* of Klymko et al. [7] for directed networks. We think that triangles play an important role in the communities' structure. In the next developments, we will try to demonstrate this assumption.

### 3.2 *Algorithm 1*

In this section, we introduce our algorithm for community detection in weighted graphs as illustrated in *Algorithm 1*.

*Algorithm 1* starts with $v$ communities, assigning to each community of $Cs$ a distinct node of $G$. Thereby, $IW$ and $WD$ are computed as defined in the previous section. Then, it finds a collection of pairwise node-disjoint triangles, which are maximal, via $findTriangles()$. Therefore, the endpoints of each triangle provided by $nodes()$ have first to be considered to form a distinct community, especially for those having a large $t^l$ value. Then, Algorithm 1 sorts the obtained set of communities $Cs$ via $sortInitialCommunities$ () taking into account the user's choice. The user can choose between a random, a decreasing $WD$ or a decreasing $IW$ order.

From there, successive steps are carried out wherein the dominant communities in $Cs$ get bigger. To do so, the adjacent communities are pairwise compared. $sortAdjacentCommunities$ () returns the set of communities adjacent to a specific community and sorts them in a decreasing $IW$ order. $intraCompare(c_g, INW_{c_g, c_h})$ returns $true$, if $INW_{c_g, c_h} \geq IW_{c_g}$ , $false$ otherwise, whereas via $interCompare(c_g, INW_{c_g, c_h})$ $true$ is returned, if $INW_{c_g, c_h} \geq \left(WD_{c_g} - IW_{c_g} - INW_{c_g, c_h}\right) * \Omega$ , where $\Omega \in [0, 0.5]$, $false$ otherwise. Then, if the condition is met, Algorithm 1 merges the compared communities while it updates $Cs$ and $IW$. Finally, $Cs$ is returned as the set of communities detected in $G$.

For better understanding of Algorithm 1, we consider the example in Figure 1, with $\Omega = 0.1$ and $userSortChoice \leftarrow IW$ . The initialization step yields to $Cs \leftarrow \{(n_i)\}^{20}$ . After $findTriangles$() and $sortInitialCommunities$() return $Cs \leftarrow \{(n_1, n_3, n_4), (n_6, n_7, n_8), (n_{13}, n_{14}, n_{16}), (n_{17}, n_{18}, n_{20}), (n_9, n_{10}, n_{11}), (n_2), (n_5), (n_{12}), (n_{19})\}$. Then, the adjacent communities are pairwise compared allowing dominant communities to gain in members. The first community considered is $(n_1, n_3, n_4)$, the latter is compared with $\{(n_2), (n_9, n_{10}, n_{11})\}$. The merging condition is met with $(n_2)$ because $intraCompare((n_1, n_3, n_4), INW_{(n_1, n_3, n_4), (n_2)}) = true$; $7 \geq 0$ and $interCompare((n_1, n_3, n_4), INW_{(n_1, n_3, n_4), (n_2)}) = true$; $7 \geq 0$, thereby $(n_1, n_3, n_4)$ and $(n_2)$ are merged. However, the merging condition is not met with the next community $(n_9, n_{10}, n_{11})$ because $intraCompare((n_1, n_2, n_3, n_4), INW_{(n_1, n_2, n_3, n_4), (n_9, n_{10}, n_{11})}) = fals$; $3 \not\geq 14$ and $intra Compare((n_9, n_{10}, n_{11}), INW_{(n_1, n_2, n_3, n_4), (n_9, n_{10}, n_{11})}) = false$; $3 \not\geq 7$ . The same steps for occurs the community $(n_6, n_7, n_8)$, $(n_{13}, n_{14}, n_{16})$ and $(n_{17}, n_{18}, n_{20})$ leading to $Cs \leftarrow \{(n_1, n_2, n_3, n_4), (n_5, n_6, n_7, n_8), (n_{13}, n_{14}, n_{15}, n_{16}), (n_{17}, n_{18}, n_{19}, n_{20}), (n_9, n_{10}, n_{11}), (n_{12})\}$. The next community to be considered is $(n_9, n_{10}, n_{11})$ . The latter is compared with $\{(n_{12}), (n_1, n_2, n_3, n_4), (n_5, n_6, n_7, n_8), (n_{13}, n_{14}, n_{15}, n_{16}), (n_{17}, n_{18}, n_{19}, n_{20})\}$. The merging condition is met with $(n_{12})$ because $intra Compare((n_9, n_{10}, n_{11}), INW_{(n_9, n_{10}, n_{11}), (n_{12})}) = true$ ; $7 \geq 7$ and $interCompare((n_9, n_{10}, n_{11}), INW_{(n_9, n_{10}, n_{11}), (n_{12})}) = true$ ; $7 \geq (3 * 0.1)$ , whereas the merging condition is not met for the remaining adjacent communities. No other community merging is possible. Thus, Algorithm 1 ends with $Cs \leftarrow \{(n_1, n_2, n_3, n_4), (n_5, n_6, n_7, n_8), (n_{13}, n_{14}, n_{15}, n_{16}), (n_9, n_{10}, n_{11}, n_{12}), (n_{17}, n_{18}, n_{19}, n_{20})\}$ which are, respectively, coloured in Figure 1 by yellow, purple, blue, grey and orange.

In order to avoid re-computing **INW** at each community comparison, an adjacency list of communities with the corresponding **INW** is memorised. This adjacency list is updated after each community merging. For instance, the adjacency list of the graph in Figure 1 is showed in (5).

$(n_1, n_2, n_3, n_4): \rightarrow \{(n_9, n_{10}, n_{11}, n_{12}): 3\}$

$(n_5, n_6, n_7, n_8): \rightarrow \{(n_9, n_{10}, n_{11}, n_{12}: 3\}$

$(n_9, n_{10}, n_{11}, n_{12}): \rightarrow \{(n_1, n_2, n_3, n_4): 3\} \rightarrow \{(n_5, n_6, n_7, n_8): 3\} \rightarrow \{(n_{17}, n_{18}, n_{19}, n_{20}): 3\}$ **(5)**

$(n_{13}, n_{14}, n_{15}, n_{16}): \rightarrow \{(n_9, n_{10}, n_{11}, n_{12}): 3\}$

$(n_{17}, n_{18}, n_{19}, n_{20}): \rightarrow \{(n_9, n_{10}, n_{11}, n_{12}: 3\}$

**Algorithm 1.**

---

**Input:** the graph $G$, $\Omega$ and $userSortChoice$.
**Output:** a collection of communities $Cs$.

    **BEGING**

    ***var*** $Cs \leftarrow \{c_g\}^v$,      $c_g \leftarrow \{n_g\}$,      $\forall g \in \{1 \dots v\}$;

    $INW_{c_g} \leftarrow 0$,      $WD_{c_g} \leftarrow WD_{n_g}$,      $\forall g \in \{1 \dots v\}$;

    ***var*** $triangles \leftarrow findTriangles(G)$;

    **foreach** $tri$ **in** $triangles$ **do**

        $Cs \leftarrow Cs \backslash nodes(tri)$;

        $Cs \leftarrow Cs \cup \{c_{|Cs|+1}\}$ where $c_{|Cs|+1} \leftarrow \{nodes(tri)\}$;

        Compute $intra\,weight_{c_{|Cs|+1}}$ and $WD_{c_{|Cs|+1}}$;

    **end foreach**

    $Cs \leftarrow sortInitialCommunities(Cs, userSortChoice)$;

    **do**

        **foreach** $c_g$ **in** $Cs$ **do**

            ***var*** $sortedCs \leftarrow sortAdjacentCommunities(c_g, Cs, G)$;

            **foreach** $c_h$ **in** $sortedCs$ **do**

                **if** (( $intraCompare(c_g, INW_{c_g,c_h})$ **and** $interCompare(c_g, INW_{c_g,c_h})$ ) **or** ($intraCompare(c_h, INW_{c_g,c_h})$ **and** $interCompare(c_h, INW_{c_g,c_h})$)) **then**

                    $c_g \leftarrow c_g \cup c_h$;

                    $IW_{c_g} \leftarrow IW_{c_g} + IW_{c_h} + INW_{c_g,c_h}$;

                    $Cs \leftarrow Cs \backslash c_h$;

                **end if**

            **end foreach**

        **end foreach**

    **while** **(**a merging is possible**)**

    **Return** $Cs$;

    **END.**

---

## 4   COMPUTATIONAL STUDY

In this section, we conduct a computational study of *Algorithm 1* on instances generated with LFR benchmark. The LFR generation scheme uses power laws for degrees and community size distributions. $\mu_t$ and $\mu_w$ are the mixing parameters which, respectively, denote the proportion between the internal and external edges for a vertex with respect to its community belonging and $WD_{n_g}$ proportion for the weighted case. For example, a specific node will have $1 - \mu\_t$ edges within its community and $\mu_t$ edges out of its communities. For further details about LFR benchmark we refer to [9], and to [10] for the weighted version. The number of vertices for the generated instances are: 1000 and 5000. The community sizes are in $[20, 100]$ for 1000 nodes and in $[20, 500]$ for 5000 nodes. The average node degree is set to 20 whereas the maximum node degree is 50 for 1000 nodes and 100 for 5000 nodes. For each $\mu_t$ and $\mu_w$ ten instances are generated. Additionally, for all the instances, we set $userSortChoice$ and $\Omega$ of $Algorithm\ 1$, respectively, to $IW$ and $0.1$.

The average values of the instances' characteristics with, respectively, 1000 and 5000 are presented in Table 1 and Table 2. Hereafter, we give some details on the experimental environment. All the tests are done on Intel Core i7 2.7 GHz processor and 8 GB of RAM, under Windows 8 OS. As an instance, Figure 2 an example is generated with the LFR benchmark where $v = 5000$ and $m \approx 47600$, depicted using $NLCOMS$ (see §5.2), where node colours represent the detected communities.

**Table 1: characteristics of the generated instances with 1000 nodes**

| $\mu_t$ - $\mu_w$ | Vertices ($v$) | Edges ($m$) | Triangles ($T$) |
|---|---|---|---|
| 0.1 - 0.1 | 1000 | 9776 | 6210,6 |
| 0.2 - 0.2 | 1000 | 9773,1 | 4534,1 |
| 0.3 - 0.3 | 1000 | 9614,7 | 3350,6 |
| 0.4 - 0.4 | 1000 | 9824,3 | 2472,9 |
| 0.5 - 0.5 | 1000 | 9790,4 | 1828,8 |
| 0.6 - 0.6 | 1000 | 9674,3 | 1225,9 |
| 0.5 - 0.1 | 1000 | 9773,7 | 1736,1 |
| 0.5 - 0.2 | 1000 | 9671,2 | 1717,3 |
| 0.5 - 0.3 | 1000 | 9840,5 | 1797,7 |
| 0.5 - 0.4 | 1000 | 9741,6 | 1764,7 |
| 0.5 - 0.5 | 1000 | 9718,3 | 1830,5 |
| 0.5 - 0.6 | 1000 | 9846,1 | 1763,5 |

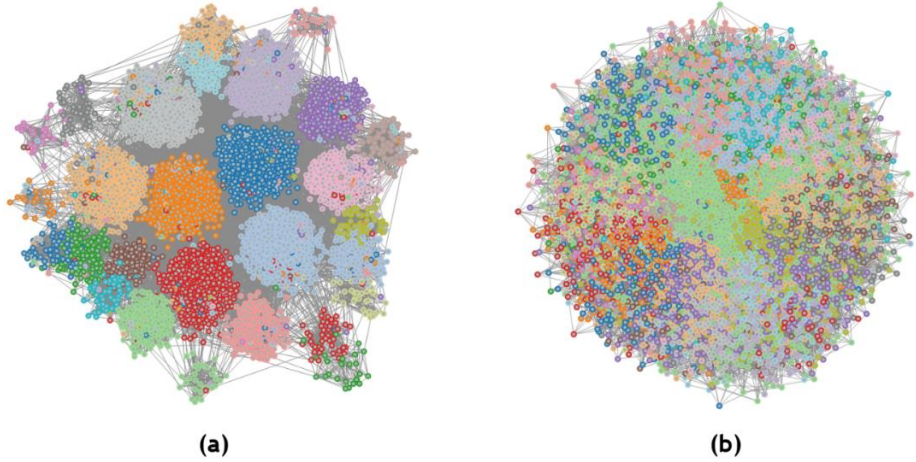

**(a)**              **(b)**

**Figure 2: LFR instances where $v = 5000$ and $m \approx 47600$, (a) $\mu_t = \mu_w = 0.1$, (b) $\mu_t = \mu_w = 0.4$.**

One have to compare the ground-truth communities of the LFR benchmark and a proposed community partition. To do so with the proposed communities of *Algorithm 1* we use *Rand Index* of [18], noted here *RI* expressed by formula (6).

$$RI(Cs_1, Cs_2) = \frac{m_{1,1} + m_{0,0}}{m_{1,1} + m_{1,0} + m_{0,1} + m_{0,0}} \qquad (6)$$

Where $m_{1,1}$, $m_{0,0}$, $m_{1,0}$ and $m_{0,1}$ represent how often any couple of nodes are respectively within a community in both partitions $Cs_1$ and $Cs_2$, in different communities in both partitions, within a community of $Cs_1$ only (hence in different communities of $Cs_2$) and in different communities of $Cs_1$ but within a community in $Cs_2$. Closer values of *RI* to 1 more *Algorithm 1* returns community partition, which matches the ground-truth in terms of nodes couple classification. Figure 3 shows the results obtained by *Algorithm 1* for 1000 nodes. The blue dots represent the case where $\mu_t = \mu_w$ whereas the red ones represent the case where $\mu_t = 0.5$. Like in [8], the second case assumes that there is as much edges within a community as for the other communities for each node, which means that there is no topological community dominance. In the latter, a community detection algorithm devised for weighted networks has to rely only on weights (i.e., $\mu_w$) to identify communities. In Figure 4, the relative completion time and the modularity are presented. Figure 5 depicts the average numbers of communities detected by *Algorithm 1* for the instances of Table 1.

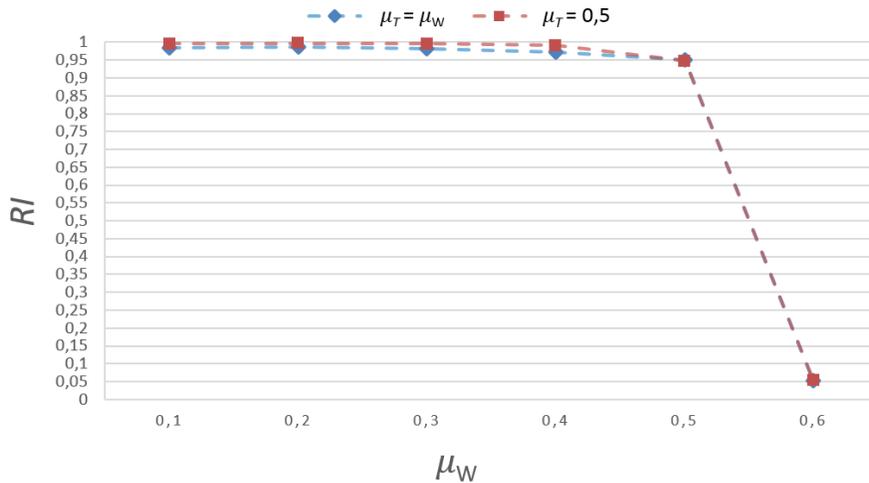

Figure 3: results of *Algorithm 1* on LFR generated instances of 1000 nodes

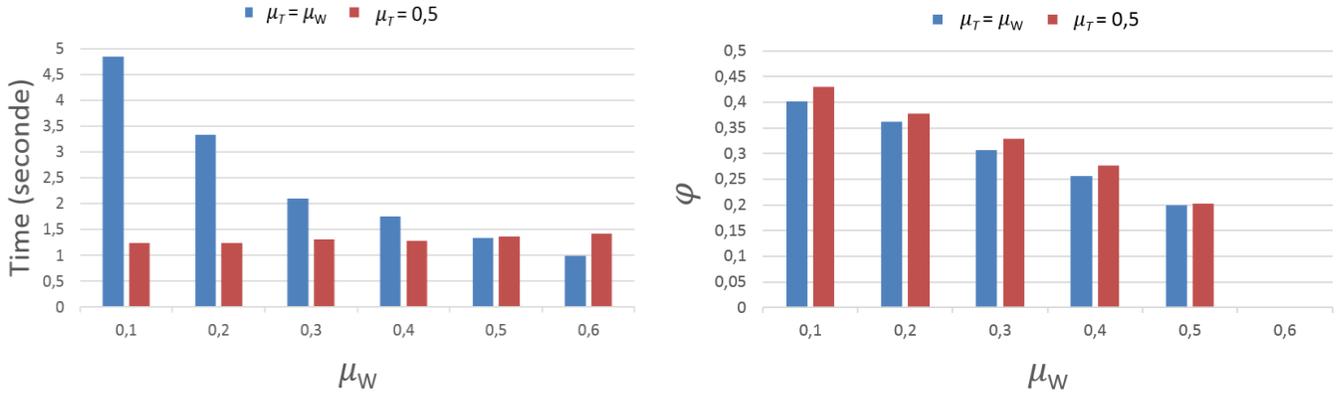

Figure 4: time and modularity of the results of Figure 3

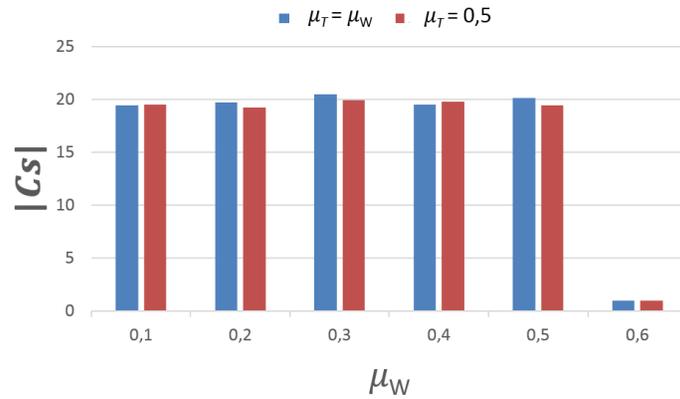

Figure 5: averages of the number of communities identified by *Algorithm 1*

From Figure 2 we remark that *Algorithm 1* has a high *RI* value up to 0.5 due to triangles which have the particularity to catch the community structure as a first step, then the merging steps add the remaining vertices to each community. After this limit value (i.e., for 0.6) only one community is detected that is because there is more inter-community edges than within communities edges. We notice also that even though there is a balance between the number of inter-community edges and within community edges *Algorithm 1* detects the communities relying only on the weights.

Figure 6 shows the results obtained by *Algorithm 1* for 5000 nodes, whereas in Figure 7, the relative completion time and the modularity are presented. Figure 8 depicts the average numbers of communities detected by *Algorithm 1* for the instances of Table 2.

**Table 2: characteristics of the generated instances with 5000 nodes**

| $\mu_t$ - $\mu_w$ | *Vertices* ($v$) | *Edges* ($m$) | *Triangles* ($T$) |
|---|---|---|---|
| 0.1 - 0.1 | 5000 | 47643 | 23334,4 |
| 0.2 - 0.2 | 5000 | 47600,3 | 16501,1 |
| 0.3 - 0.3 | 5000 | 47556,2 | 11697,9 |
| 0.4 - 0.4 | 5000 | 47467,7 | 8858,9 |

| 0.5 - 0.5 | 5000 | 47346 | 5844,6 |
|---|---|---|---|
| 0.1 - 0.1 | 5000 | 47643 | 23334,4 |
| 0.5 - 0.1 | 5000 | 47954,1 | 5983 |
| 0.5 - 0.2 | 5000 | 47576,7 | 5930,7 |
| 0.5 - 0.3 | 5000 | 47854,4 | 5704,5 |
| 0.5 - 0.4 | 5000 | 47533,8 | 5481,9 |
| 0.5 - 0.5 | 5000 | 47863,6 | 5768,4 |

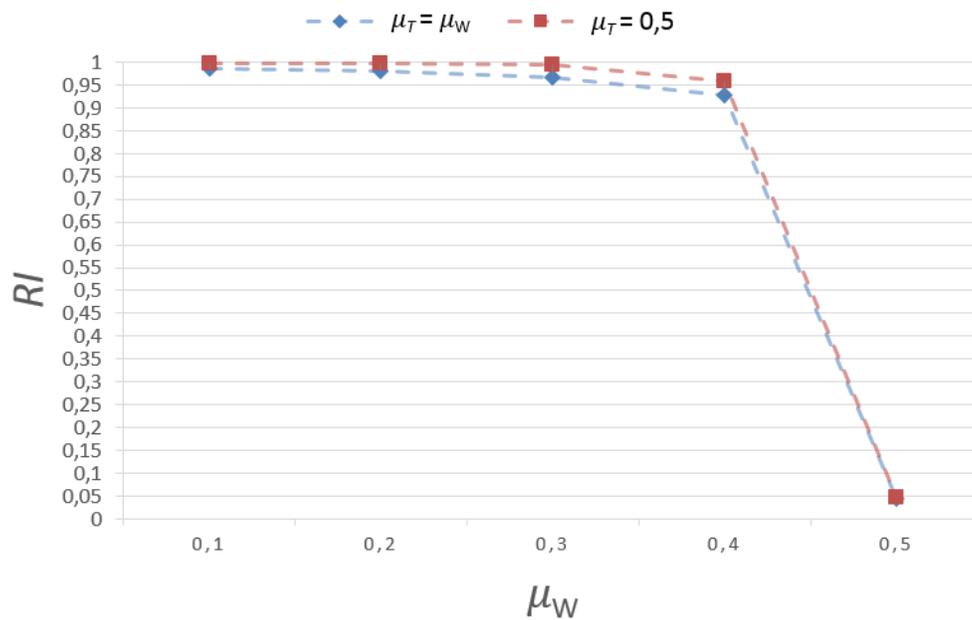

Figure 6: results of *Algorithm 1* on LFR generated instances of 5000 nodes

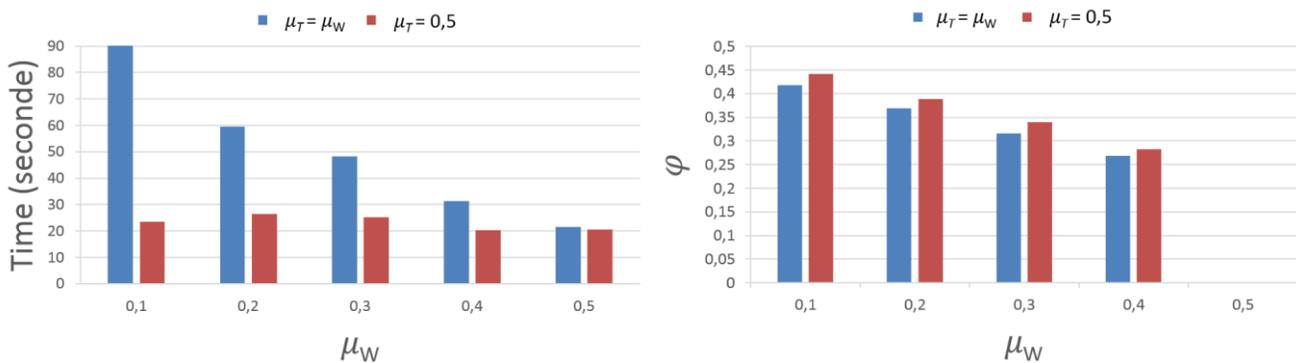

Figure 7: time and modularity of the results of Figure 5

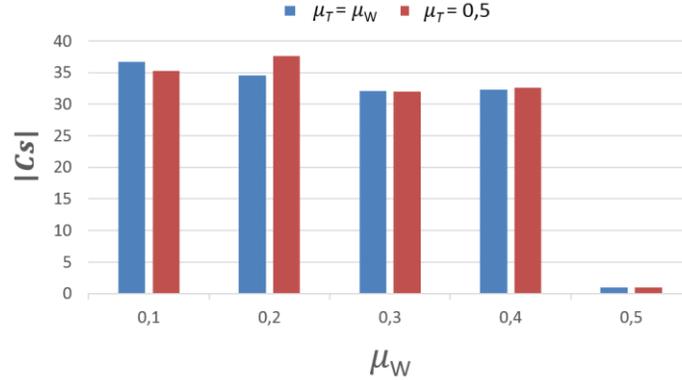

Figure 8: averages of the number of communities identified by *Algorithm 1*

With these additionally instances of Table 2, Figure 6 shows that *Algorithm 1* still have high *RI* values. However, the limit becomes equal to 0.5 which remains an acceptable limit. We point out also that *Algorithm 1* performs slightly better when $\mu_t = 0.5$ and $\mu_w$ varies, this observation holds also in Figure 2. Regarding the completion time we observe that it decreases when the number of triangles in the graph decreases.

## 5   SOCIAL NETWORK ANALYSIS

### 5.1   Info-RSN project

In order to assess *Algorithm 1* on real-world application, we focus on community detection in social networks and we consider the ANR Info-RSN project. In the latter, 17 millions of tweets are provided and the objectives are to provide an explanation on how the information is shared through social network (Twitter) and also to detect communities within Twitter. To fulfil this need, we have done pre-processing step on the Info-RSN database in order to get a graph model in which nodes are the persons who tweet and the edges represent either re-tweet or mention which represent, respectively, person B re-tweet the tweet of person A and person A is mentioned by person B in its tweet. Thereby, the communities induced by re-tweet edges and those induced by mention edges are complementary to investigate the sharing and propagation phenomena.

### 5.2   *NLCOMS*

An application is created to visualize and to interpret the network and the detected communities. *NLCOMS*, for (Node-Link and COMmunitieS), is a visual interactive application for social network analysis based on node-link representation of graphs and circle packing for the detected communities. Figure 9 gives a global sight of *NLCOMS* and Figure 10 shows the circle packing representation of communities detected by *Algorithm 1* for the graph in Figure 2 (a). The circles size is proportional to the node graph degrees.

In Figure 9, the main view is divided into two panels labelled Node-Link and Communities. The first is the area where the node-link representation is depicted whereas the detected communities are depicted in the second panel. *NLCOMS* proposes other panels labelled Node-link attributes, Community detection, Data filters, Layout parameters and Stats views, in which the user can, respectively, enter the desired number of nodes or filter the graph relatively to the graph degree (i.e., we keep each connected component containing a node having a graph degree value greater than or equal to the desired user graph degree value), set the value of $userSortChoice$ used in $sortInitialCommunities()$ of *Algorithm 1*, filter the raw data of the Info-RSN database, hide labels or change the edge length and rely on some performed statistics on a selected community. *NLCOMS* permits to the user to decide which relationship type are represented by the edges. In this context, the edge weight $e^w \in E^w$ represents the number of re-tweets or mentions between two persons. *NLCOMS* allows also tweets filtering relative to a particular tweet theme, a specific media or the date of publication.

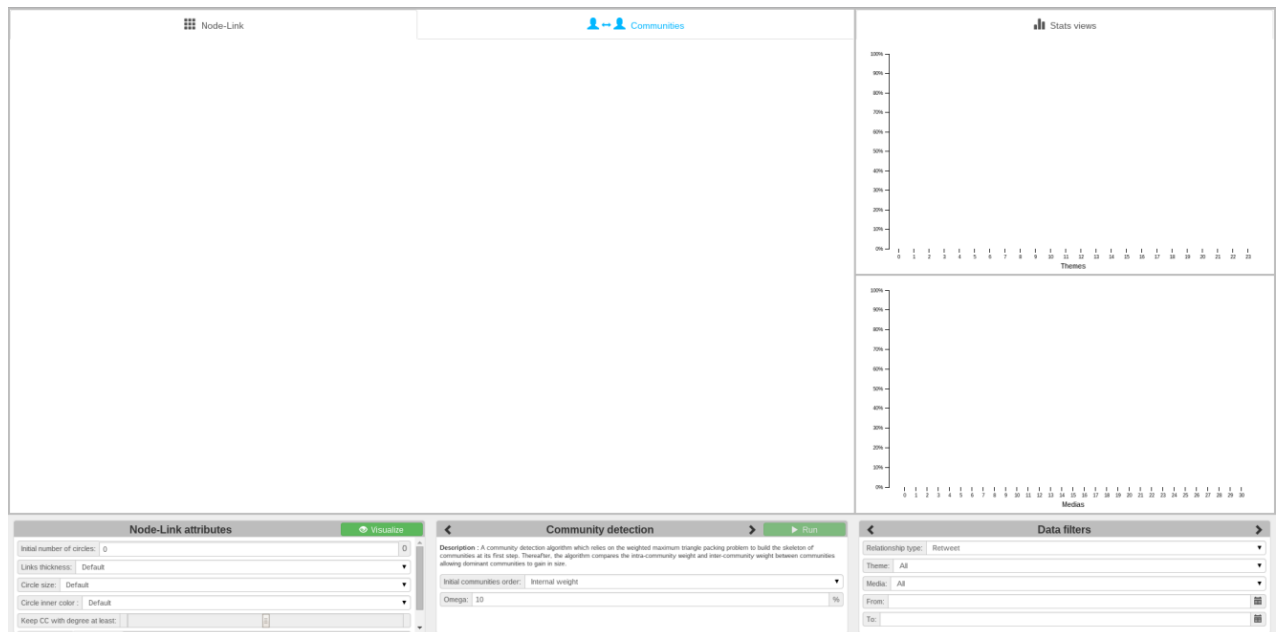

Figure 9: global sight of *NLCOMS*

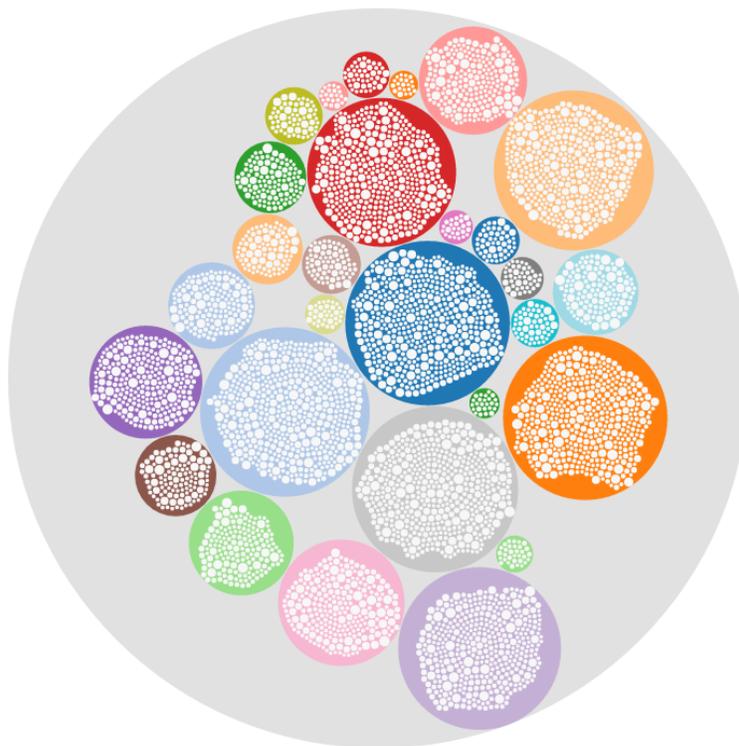

Figure 10: Circle packing representation using *NLCOMS* for the graph in Figure 2 (a)

Besides, we use the visual variable of Bertin [2] on the node size, the node shape, the edge thickness and the lightness of the inner node colour which visually represent, respectively, number of followers of the person, whether the person who tweet uses twitter as reporter or as ordinary user, number of re-tweets or mentions between two persons, number of tweets of the person. Additionally, the node shape outline is coloured relative to the community node's membership. Thus, each community is coloured by a colour. Moreover, complementary and interactive bar charts help the expert user to build his/her own ideas about communities' characteristics and give some statistics about thematic and Medias source proportions within a selected community. Additionally, circular progress bar tells the expert user the tweets proportion for each community member of its tweets with respect to a selected thematic or Media source in the community representation.

*NLCOMS* is driven by the visual information-Seeking Shneiderman's Mantara "*Overview First, Zoom and Filter, Then Details-on Demand*" [21]. A global view gives to the expert user a sight of the networks structure, after a zoom on the network or on the detected communities is provided. Details appear when the expert user selects a community or a member. Effortless interactions (i.e., right and left mouse click) are utilized avoiding pointless extra cognitive load. The steps of the interactive visualization on which *NLCOMS* relies are presented in Figure 11.

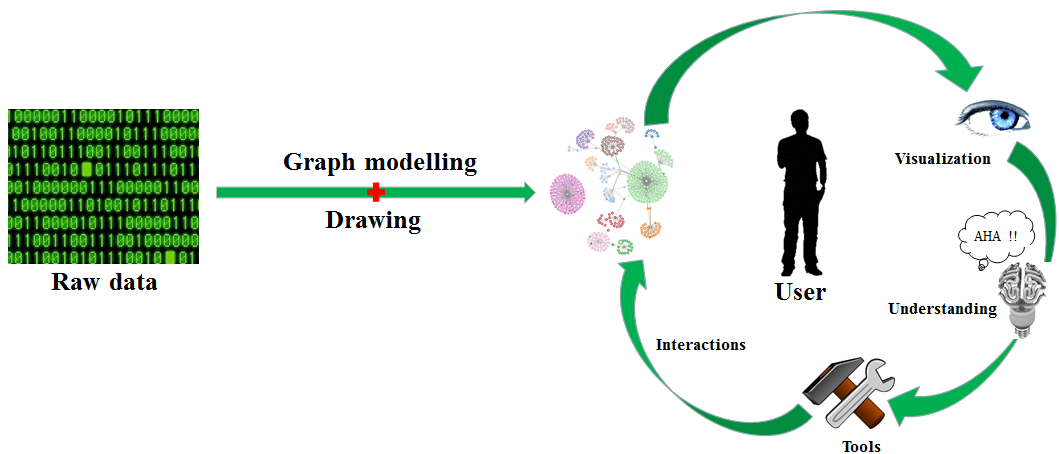

**Figure 11: interactive visualization steps**

## 5.3 Data sets and results

Representative samples from the Info-RSN database are considered. Table 3 illustrates the samples' characteristics of the Info-RSN database. Like for Section 4, we set $userSortChoice$ and $\Omega$ of *Algorithm 1*, respectively, to $IW$ and $0.1$.

**Table 3: Samples' graphs characteristics from the Info-RSN database**

| Data sets | $Initial\ (v,m)$ | $Filtred\ (v,m)$ | Thematic(s) | Medias source | Edge type |
|---|---|---|---|---|---|
| Data set 1 | (1000, 935) | (554, 556) | all | all | re-tweet |
| Data set 2 | (2000, 2193) | (2000, 2193) | all | lefigaro.fr | re-tweet |
| Data set 3 | (3000, 2999) | (3000, 2999) | war | all | re-tweet |
| Data set 4 | (1000, 1027) | (1000, 1027) | all | liberation.fr | mention |

In order to understand how the information is shared and the underlying semantic on the networks, the following set of visual tasks are considered:

- What are the mains actors in the network? Have they followers?
- There is atypical behaviour? Who uses Twitter as reporter?
- What are the structures of the communities?
- What are the mains thematics addressed and the main media source within a community?
- How often a specific member tweet in a specific thematic or from a specific media source?
- Is there a structural difference between networks with re-tweet edge type and mention edge type?

### 5.3.1 Data set 1 for edge re-tweet type analysis

In data set 1 the network is sparse with communities having star-like shape. From Figure 12, for the main actors we can easily distinguish their followers in the circle packing representation or in the node-link representation. The node size in the circle packing representation depicts the number of followers of a community member. Furthermore, we notice a bridge-like behaviour, which links the centroid of two star-like communities. One would say that the centric member emits the original tweet and the followers propagate them. For example, in Figure 13, if the expert user selects the pink community bar charts appear, allowing theme or media selection. The height represents the tweet percentage classified as tweets dealing with the selection theme (media) with respect to the total number of tweet within the community. The lightness bar colour depicts the proportion between community members that tweet at least one time in the selected thematic (media) and the members within the community which means darker are the bar more community members belong to it. In this context the theme 'sport' and the media 'slate.fr' are selected in Figure 13.

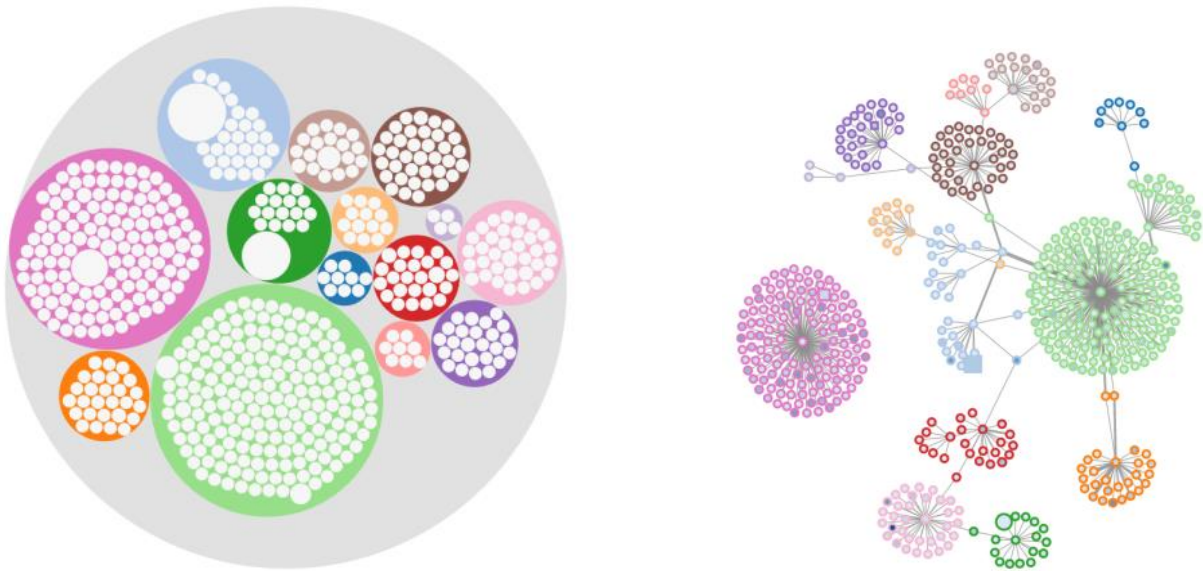

**Figure 12: Data set 1**

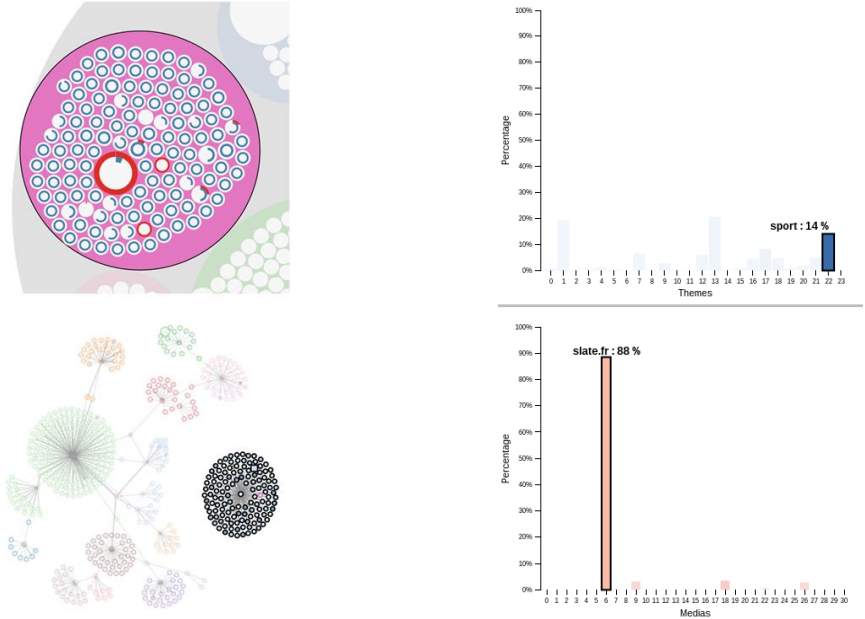

**Figure 13: Selection of the pink community in Figure 12**

From Figure 13 we observe that 88% of tweets in the pink community tweet are classified in the media 'slate.fr' and 14% talks about 'sport'. Additionally, the totality of the tweets of the main actor (i.e., having the most of followers depicted by the bigger circle in the pink community) is from 'slate.fr'. The combination of these representations enhances the understanding of community structure while grasping gradually and interactively the underlying information.

### 5.3.2 Data set 2 for re-tweet trends

In data set 2 the network is less messy, the communities are more distinguishable. Unlike in Figure 12, in Figure 14 the circle size is proportional to the number of tweets of each community member. *NLCOMS* provides circle ordering with respect to their size, which allows answering quickly to the following question; how is the most (less) active member?

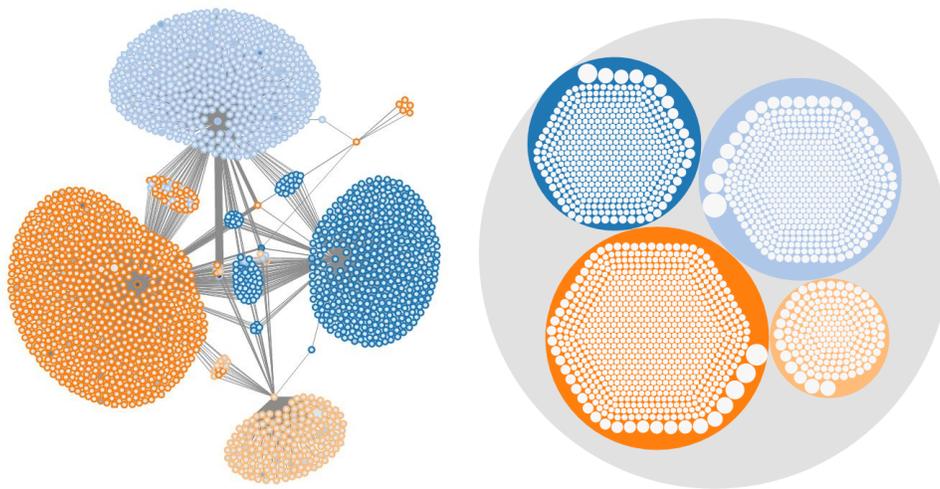

**Figure 14: Data set 2**

The thematic or media source filtering is done only on the initial tweets, which means that if a re-tweet occurs but the thematic of the re-tweet differs from the user expert filter, we have decided to keep the couple (tweet, re-tweet) in the network. The motivation behind this is to analyse the re-tweet trends wither the thematic or the media source change.

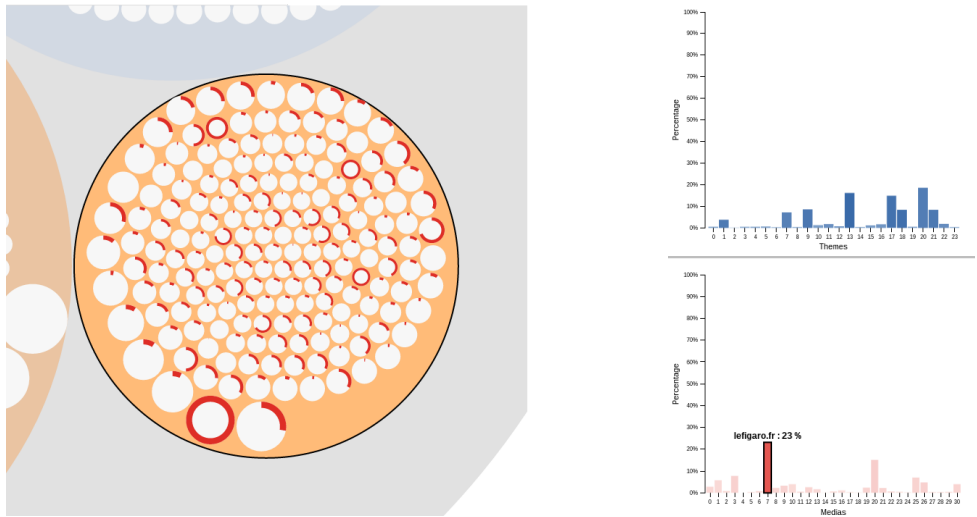

**Figure 15: Selection of the light orange community and the 'lefigaro.fr' bar**

As an example, in Figure 15, after selecting the light orange community, the histograms give the information that the dominant media in the community is 'lefigaro.fr', which reflects the initial filter. Additionally, we remark that some members tweets only from 'lefigaro.fr' whereas others have a great tweet thematic ratio. By taking these observations into account the expert user can infer the main active members on a specific media (theme).

### 5.3.3  Data set 3 for cross thematics

In Figure 16, data set 3 shows that there are some thematics, which are cross connected. For example, the thematic filter is 'war' but the main and the most tweeted thematic is 'international'.

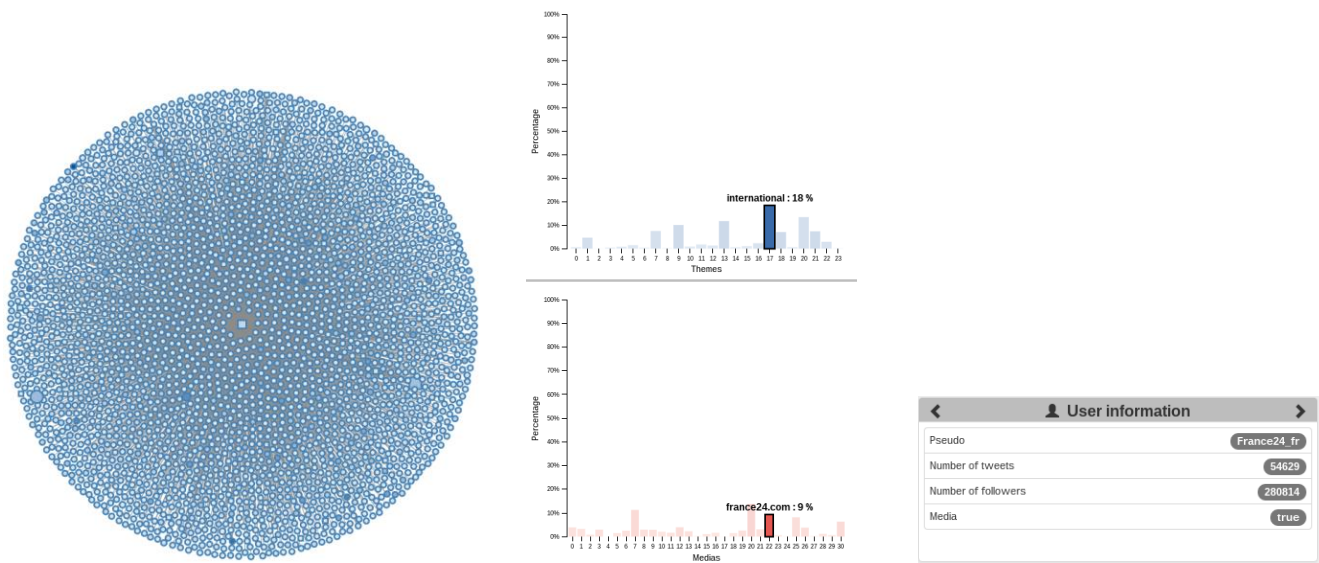

**Figure 16: Data set 3 and the user information of the centroid member**

Additionally, the centroid member of the unique community is a reporter for 'france24', which is a breaking news TV channel and all other members are connected to the aforementioned reporter, suggesting that this member is the main actor in the network.

### 5.3.4 Data set 4 for edge mention type analysis

In data set 4, we investigate the mention edge type; the sample is illustrated in Figure 17. As for the previous data sets the used filter (i.e., selection of 'liberation.fr' as media) is dominant in the light blue community. An interesting observation is that the majority of the bridge-like behaviours are reporters represented by squares.

Figure 18 presents completion times and the modularity of *Algorithm 1* for the Info-RSN data sets.

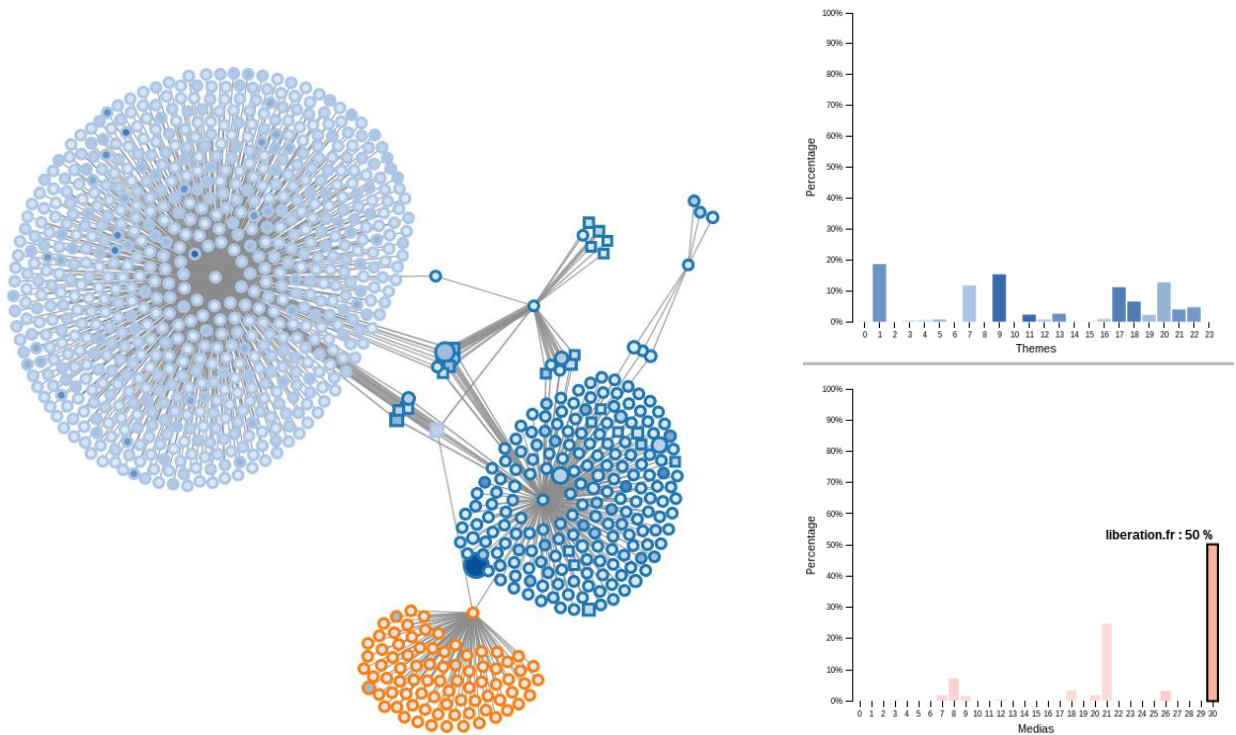

**Figure 17: Data set 4 where the light blue community is selected**

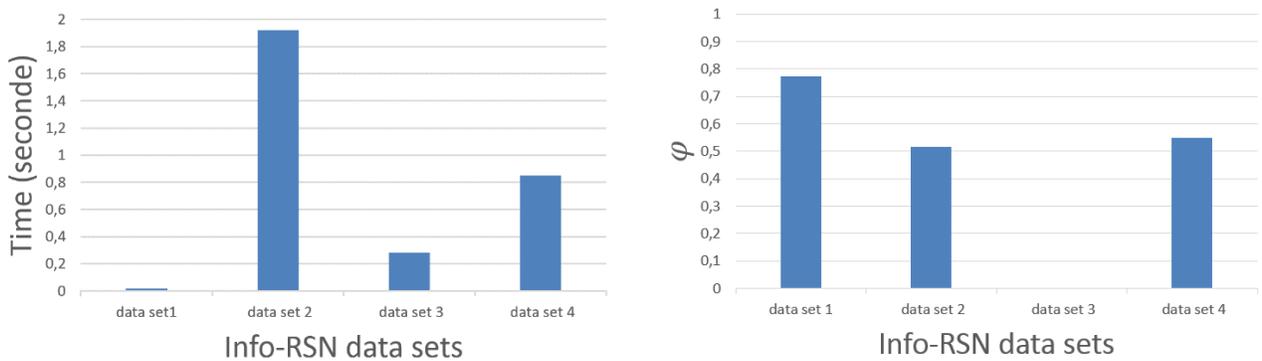

**Figure 18: Completion times and the modularity of *Algorithm 1* for the Info-RSN data sets**

## 6  CONCLUSION

In this work, we present our algorithm for community detection in networks. *Algorithm 1* uses the triangles obtained by a feasible solution of the weighted MTP problem as starting point for community detection. Then, communities are compared allowing dominating communities to gain in size. The proposed algorithm is tested on the LFR benchmark. The results show that *Algorithm 1* has high *RI* up to $\mu_w = 0.5$ for 1000 vertices and $\mu_w = 0.4$ for 5000 vertices, even when $\mu_t = 0.5$ and $\mu_w$ varies. Furthermore, we propose an interactive visualization approach relying on *NLCOMS* to reveal the structure and the underlying semantic within communities. Additionally, one among the objectives of this study is to understand how the information is shared on social network, especially on Twitter. To fulfill this need, we investigate real-world data of the info-RSN ANR project.

As perspectives, it is worthwhile to consider the dynamic context where the aim is not only to distinguish the interconnected communities at each time-point but also to devise an algorithm which takes advantage from the previous time-points (Nguyen et al. [15]), and the related visualization which permits to depict the evolution of communities' structure (Reda et al. [19]).